# RIS-aided multi-hop backhauling for 5G/6G UAV-assisted access points

Salim Janji, Paweł Sroka
Poznan University of Technology, Institute of Radiocommunications
salim.janji@doctorate.put.poznan.pl, pawel.sroka@put.poznan.pl

*Abstract*—Drones are envisaged as an important part of the future 6G systems. With the possibility of their fast deployment they provide additional connectivity options in form of a hotspot. However, typically in such a use case they require provisioning of a wireless backhaul link to facilitate their proper operation, which might be a challenging task in urban environment. One of the possible ways to connect such nodes is to use the integrate access and backhaul (IAB) approach, where part of the spectrum dedicated for user access at the base station is used for wireless backhauling. Thus, in this work we consider the problem of establishing a multi-hop wireless backhaul link following the IAB concept, with the aid of drone relay stations (DRSs) and reconfigurable intelligent surfaces (RISs). We formulate the problem of coverage improvement with fixed number of relays assuming certain throughput requirements on the backhaul. We show with simulations that the use of RISs allows for improvement of coverage in such a scenario or reduction in the number of involved nodes to provide the required backhaul.

## I. INTRODUCTION

ONE of the key use cases considered with 5G and beyond wireless networks is the provisioning of enhanced mobile broadband (eMBB) services through deployment of a large number of small base stations (BSs), creating the so-called ultra-dense networks (UDN). One of the challenges when deploying UDN is provisioning of backhaul network that will connect the numerous nodes to the core network [1]. Solutions relying on wired backhauling might be not available, congested or providing only limited capacity. Thus, fast and economic deployment of backhaul infrastructures providing the required capacity is considered a key enabler for UDN.

Wireless backhaul provisioning with the relay nodes is considered in 5G system, among other solutions, in the framework of integrated access and backhaul (IAB) concept. The idea behind IAB is to serve simultaneously end users and relay nodes over wireless links with a single macro BS for coverage extension. Two distinct approaches can be considered with IAB: in-band backhauling, where the same frequency resources are used for user access and backhauling, resulting in possible interference between these links, and out-of-band backhauling, where separate frequency resources are used for user access and backhaul provisioning, thus reducing the interference problem [2].

Among the emerging technologies introduced in 5G systems, unmanned aerial vehicles (UAVs) are envisaged as an important part of wireless access networks, providing, among others, reconfigurable on-demand access or relay nodes. The main advantage of such UAV-assisted wireless communications is its low deployment cost, suitable for emergency scenarios, or temporary network capacity boosts. Furthermore, the wireless channels between UAVs and ground nodes typically is a line-of-sight (LoS) link, thus providing better wireless coverage and higher communication throughput. With the fully controllable mobility of UAVs in three-dimensional (3D) airspace, they can adaptively change their locations to improve the communications performance. Therefore, one of the envisaged applications of UAVs is to provide wireless backhaul connectivity to the small BSs [3]. In a scenario when multiple UAVs are deployed as quasi-stationary aerial relay nodes, by optimizing their location and the routing path, a wireless multi-hop backhaul link can be established between a small BS and a macro BS.

The deployment and control of UAV-enriched wireless multi-hop network is a challenging task. UAV size, mobility, payload, energy consumption and the related battery endurance are the key constraints, and are subject to performance degradation, making it difficult to incorporate a drone as a reliable node for 5G or beyond wireless network. Additionally, in an urban environment, UAVs may experience LoS link blockage due to high rise buildings, thus requiring use of a high number of such nodes to provide high capacity backhaul. In order to overcome these problems, advanced transmission techniques making use of reconfigurable intelligent surfaces (RIS) can be considered to improve the capacity or reliability of wireless links [4] in such a challenging scenario, where establishing LoS links between UAVs (or UAV and BS) is not possible or requires increase in the number of operating drones.

RISs, also called intelligent reflecting surfaces (IRSs) or large intelligent surfaces (LISs), are arrays with a large number of reflecting elements that can be used to change the amplitude, frequency or phase of the incident signals [5]. RIS are capable of mitigating a wide range of challenges encountered in diverse wireless networks, proactively modifying the wireless communication environment, thus providing gains in capacity, reliability, sustainability, coverage, and secrecy performance of wireless communication. One of the main advantages of RISs is the ease of their deployment. Made of nearly-passive devices of electromagnetic material, they can be mounted on several structures, including building facades, indoor walls, aerial platforms, roadside billboards, etc. Furthermore, they are more energy efficient than the conventional relays, as the phase, absorption, reflection, or refraction of IRS passive reflecting elements can control the incident signals without any need of RF chains. With their capability of reconfiguring the wireless propagation environment by compensating for the power loss over long distances, they can help in forming virtual LoS links between the two transmission endpoints without direct LoS propagation. Finally, RISs are compatible with current radio technologies, supporting full-duplex and full-band transmission due to the fact that they only reflect the EM waves. Thus, their integration in existing wireless systems is possible without significant changes in hardware.

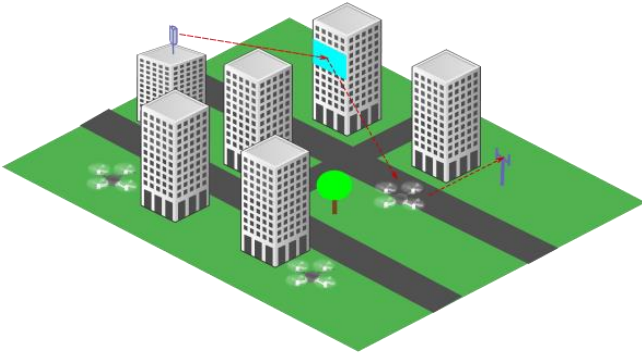

Fig. 1. Illustration of the envisaged application scenario, where RIS and UAVs are used to provide a multi-hop backhaul link between the macro BS and a small BS.

When it comes to the use of RISs with UAVs, due to their uniform spatial configuration when installed on high altitude buildings, these can provide shorter LoS paths. Especially in urban scenarios, RISs can be employed to overcome the signal blocking by high-rise buildings when communicating with UAVs operating at low altitude. Furthermore, with RIS mounted on a mobile UAV rather than a fixed wall or building, gains in coverage and flexibility of deployment are envisaged. RIS-assisted UAVs can be used to reduce the channel complexity and mitigate interference for wireless communication as well as access delay and energy consumption for UAVs [6]. By joint optimization of UAV's trajectory and resource allocation, accounting for the presence of RISs, one can achieve significant reduction in power consumption, which is one of the crucial factors in UAV-assisted wireless networks.

In this work, following the research described in [7], we consider the problem of placing the drone-based relay station (DRSs) and multi-hop transmission path selection to ensure backhaul connectivity to a given point (small BS) between obstacles, as shown in Fig. 1. Assuming the backhaul link is realized with radio communications in the framework of the 5G system, and specific requirements on throughput provided to the end node are formulated, we extend the work by adding RISs deployed in different configurations on buildings' facades. Following an assumption on the throughput requirement, we show that with RISs it is possible to increase the coverage with the use of a fixed set of UAVs, or reduce the number of drones required for backhauling for a certain served small cell.

The remaining part of this paper is structured as follows. Section II summarizes selected works on the topic of UAV positioning aiming at optimization of wireless backhaul provisioning. Section III introduces the considered system model and parameters, as well as the basic formulation of the optimization goal. Section IV presents the envisaged solution to the introduced problem along with its exemplary analysis via simulations. The following Section V outlines the possible further development of the optimization problem, and other considerations or assumptions on the investigated scenario. Finally, Section VI concludes the work.

## II. RELATED WORK

The idea of wireless backhauling with the use of UAVs gained a lot of attention in research recently, following the introduction of aerial network part concept for 6G. While there are many works focusing on the resource allocation problem for wireless backhauling with drones, only several of them include UAV positioning as one of the optimized parameters. The work presented in [8] presents the problem of resource allocation in an in-band IAB scenario where drones are used as access points using wireless backhaul to the macro BS. Power allocation as well as UAVs locations are optimized to maximize the network performance in terms of sum rate, however, only single-hop backhauling is considered. Similarly, [9] considers a similar IAB scenario, however using out-of-band backhauling in the mmWave band and aiming at optimizing the drones' locations. More sophisticated optimization problem is considered in [10], where joint UAV location, user scheduling and association and spectrum resource allocation is considered with single-hop UAV-aided wireless backhauling. Similar optimization problem is also considered in [11] applied to the cognitive radio network, thus also accounting for the inter-system interference. However, all these works do not consider a multi-hop backhaul configuration. Such problem is investigated in [12], where aspects of throughput maximization with proper UAV positioning and bandwidth and power allocation is considered. Multi-hop networking is also considered in [13], where a game-theoretic framework is proposed for backhaul optimization.

The introduction of RIS concept opened new possibilities in establishing a wireless backhaul link for UAVs. However, as the RIS idea is relatively new, only recent works consider its application with drone-assisted backhauling. In [14] a high-altitude platform system mounted RIS is considered for backhaul provisioning to drone BSs, focusing on energy-efficiency aspects. Placement and array partitioning strategies for airborne RIS are investigated, as well as optimization of phasing of array elements. Similarly, drone-mounted RIS is considered in [15], where a multi-armed bandit problem is formulated for a mmWave backhauling scenario. However, neither of these works consider a hybrid scenario with mixture of RIS and DRSs.

## III. SYSTEM MODEL AND PROBLEM FORMULATION

### A. System model

In this work we consider UAV-assisted multi-hop backhauling in an urban environment, where a single macro BS is responsible for providing coverage over the whole area with the aid of drones. We consider an out-of-band IAB scenario, where a dedicated part ($B$) of a mmWave band is reserved for backhauling purposes. We assume that there are at most $N$ UAVs available that can serve as an access point or as a relay node for multi-hop backhaul provisioning. We consider an extended Madrid grid [16] scenario, where the macro BS is located in the center of the considered area shown in Fig.2. Drones operate in a semi-static way once being deployed in any of the locations on the road intersections (marked with red stars in Fig. 2). Furthermore, we consider availability of $R$ RISs mounted on the facades of selected buildings, which can be used to increase the connectivity range of a single-hop transmission.

Due to the capacity constraints of the backhaul links, we assume that certain throughput needs to be achieved between the macro BS and UAV or between two UAVs for a link to be considered as available, thus considering the LOS links. We consider the losses due to reflections on obstacles are too high with the considered frequencies to provide enough capacity for a single hop. However, we account for the possibility of reflected transmission via RIS, where the received power level depends on the joint attenuation of the two paths: transmitter to RIS and RIS to the receiver. The link budget is then calculated based on the path-loss estimated following the formulas given below. In the case of a direct mmWave link between two stations (i.e., no RIS involved), the path loss is similar to the free-space path loss, and it is given by [17]:

$$PL_{direct}{}^{dB}(d) = PL(d_0) + 10\alpha \cdot log(d) \quad (1)$$

Where $d$ is the distance, $PL(d_0)$ is the free-space path loss at distance $d_0 = 5m$, and $\alpha$ is the path loss exponent. In our work we assume $PL(d_0) = 39$ dB for 38 GHz transmission and $\alpha = 2.13$ [9, Table 3]. As for the path loss when RIS is involved, we use the same model as in [18]:

$$PL_{RIS}{}^{dB}(d_{1\to R}, d_{R\to 2}) = PL(d_0) + 10\beta \cdot log(M^2(d_{1\to R} + d_{R\to 2})) - g_{bf} \quad (2)$$

where $M$ is the number of meta-surfaces per RIS, $\beta$ is the path loss exponent, and finally $d_{1\to R}$ and $d_{R\to 2}$ are the transmitter-RIS and RIS-receiver distances, respectively. Furthermore, we assume that the RIS is actively reflecting signals with beamforming capabilities which is accounted for by adding the gain term, $g_{bf}$, in the formula which reduces the resulting path loss. We assume that $M = 3$ and $\beta = \alpha = 2.13$.

The average link throughput can be calculated using the modified Shannon formula [19]:

$$C_i = \eta \cdot B^{(eff)} \log_2(1 + SNR_i), \quad (3)$$

where $\eta$ is the throughput efficiency of the system (fraction of data bits in total number of bits transmitted), $B^{(eff)}$ is the used effective total bandwidth and $SNR_i$ is the average signal-to-noise ratio of the $i$-th hop, calculated as:

$$SNR_i = \frac{P_i^{(TX)} g_i}{\sigma^2}, \quad (4)$$

with $P_i^{(TX)}$ and $g_i$ being the transmit power (constrained as $P_i^{(TX)} \leq P_{max}$) and the channel gain of the $i$-th link, respectively, and $\sigma^2$ representing the noise component. Channel gain can be calculated accounting for the path-loss and antenna gains of the link as $g_i = G_i^{(Tx)} G_i^{(Rx)} 10 \log_{10} PL_i^{dB}$, where $G_i^{(Tx)}$ and $G_i^{(Rx)}$ are the transmit and receive antenna gains, respectively, and $PL_i^{dB}$ is calculated using (1) or (2) for the direct or RIS-aided links, respectively. Furthermore, we assume that $\eta$ and $B^{(eff)}$ can be calculated following the 5G system specification [20].

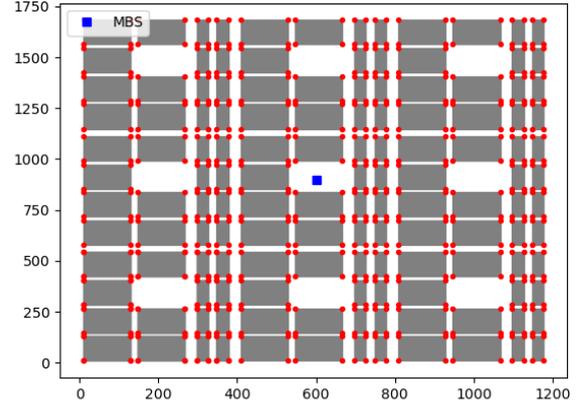

Fig. 2. A schematic of buildings layout, macro BS (MBS, marked by a blue square), and possible relaying UAVs locations (marked with red circles) for the considered Madrid grid urban environment setup.

*B. Problem formulation*

The aim of this work is to maximize the coverage of the macro BS providing backhaul opportunities capable of achieving a minimal throughput $C_{min}$, with the aid of at most $N$ UAVs operating as access points or relay nodes. Each UAV might be located in one of the predefined locations, where the drone location $L_n = (x_n, y_n)$ with $(x_n, y_n)$ being the Cartesian coordinates. Thus, the optimization problem can be formulated as:

$$\max_{\{L_1, L_2, \ldots, L_n\}} |A|, n \leq N, \quad (5)$$

where $A = \{L_k, \forall k: Q_k = 1\}$ is the set of all possible locations of an access point UAV where backhaul connectivity to macro BS can be provided. $Q_k$ is the connectivity indicator of access point location $L_k$ defined as:

$$Q_k = \prod_{i=1}^{n} q_i, n \leq N, \quad (6)$$

where $q_i$ denotes the availability of connection via the $i$-th hop of the backhaul link formulated as:

$$q_i = \begin{cases} 1 \; if \; C_i \geq C_{min} \\ 0 \; otherwise \end{cases}, \quad (7)$$

Using (3), the value of $C_i$ is calculated for the $i$-th link and it can be translated to an SNR requirement of $SNR_{min}$.

When RIS is included in communication on one of the hops, this particular link is still considered as a single-hop one, despite the RIS supporting the transmission. In such a case the formula for throughput calculation accounts for the joint path-loss of two paths in communication with RIS: Tx→RIS and reflected path RIS→Rx. Thus, equation (2) is used in channel gain estimation.

IV. SOLUTION AND EXEMPLARY EVALUATION

*A. Proposed solution including RIS availability*

Given an environment with obstacles (e.g., similar to Fig. 2) and any two points in the area, we can find if these two points are visible to each other (i.e., have a clear LOS path) using the algorithm described in [7]. There, Lee's visibility graph algorithm was applied to a graph $G(V, E)$ that contains the set of all buildings' edges, $E$, and the set of their corner vertices, $V$, which are considered the set of points for which the visibility is calculated. That is, for every point $v_i \in V$ we find the set of visible points from the set of the remaining vertices $\{v_j \in V - v_i\}$. Then, given the resulting visibility graph $V_g$ that contains edges between vertices that are visible to each other, we can apply Dijkstra's algorithm to select the shortest path through successive LOS hops.

In this work we extend the algorithm to account for new possible paths resulting from an RIS installed on a building façade. Using Lee's algorithm, we are also able to find all visible points from a given RIS. If two points are visible to the RIS then this indicates that a link can be made between them with no extra DRS involved. To proceed further, for every installed RIS we find the set of visible points from that RIS. Then, we add an edge to the earlier visibility graph for every possible pair of points.

Now that we have made the points that can communicate through an RIS visible to each other, what remains is to apply Dijkstra's algorithm to find a backhaul path to any given point. For finding the path, we define different costs for direct edges between two points and for indirect edges that go through an RIS, while also taking into consideration the expected rate of any edge, $C_{edge}$, which is obtained using (3). These costs are obtained using:

$$D_{edge} = \begin{cases} PL_{direct}^{dB} + P \text{ if direct edge and } C_{edge} > C_{min} \\ PL_{RIS}^{dB} + P \text{ if RIS edge and } C_{edge} > C_{min} \\ \infty \text{ otherwise} \end{cases} \quad (8)$$

where $P$ is a fixed penalty per hop to guide the search towards paths with least hops, and $\infty$ is a sufficiently large number indicating that the edge does not satisfy the rate requirement and therefore it should be avoided if possible.

### B. Simulations

In our simulations we compare two scenarios: the case where there are no RISs; and the case where two RISs are installed in the middle square. These RISs provide virtual LOS paths between any two points that are visible to the RIS within the obstacles, the channel through the RIS also includes a beamforming gain of $g_{bf}$ as previously mentioned. These points belong to the set of feasible DRS locations which is obtained from the buildings' corners. Table 1 lists the simulation parameters assumed.

*Table 1: Assumed simulation parameters.*

| Parameter | Value |
|---|---|
| Transmission power $P^{(Tx)}$ | 100 mW |
| RIS beamforming gain $g_{bf}$ | 15 dB |
| Noise power $\sigma^2$ | -131 dBm |
| Throughput efficiency $\eta$ | 0.82 |
| Effective bandwidth $B^{(eff)}$ | 18.72 MHz |
| Required SNR: $SNR_{min}$ (Corresponding to minimum throughput $C_{min}$) | 41, 31, 21, and 11 dB (200, 150, 100, and 55 Mbps) |

The first results we show in Fig. 3 indicate the required number of DRS hops for providing connectivity to each point from the considered set of points. In this case we assume $SNR_{min} = 31$ dB and no RISs installed. We compare these results with the case where $R=2$ RISs are installed (Fig. 4) that clearly shows the increased reachability of the MBS to farther destinations. Furthermore, less hops are required for reaching the same earlier destinations marked in Fig. 3 and the whole map can be covered with 7 hops only, whereas 8 hops were needed for covering almost the entire map without RISs as shown in Fig. 3.

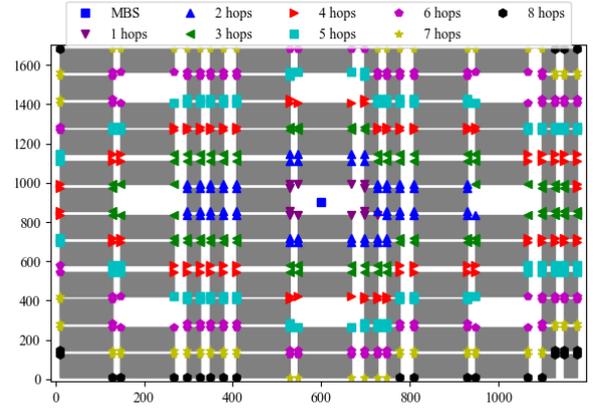

Fig. 3. Required number of DRS hops to reach each point is indicated by the marker color and shape. Notice that not all map is reachable with a maximum of 8 DRS hops.

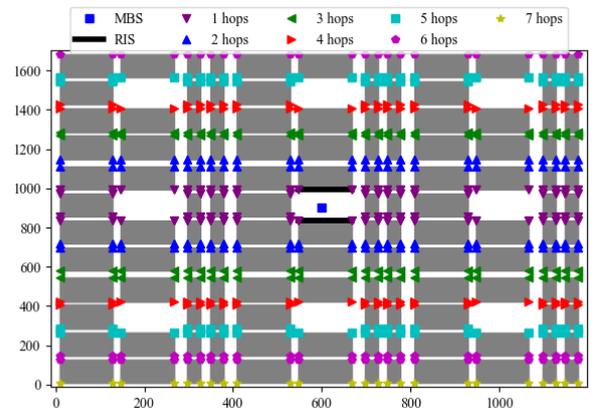

Fig. 4. Required number of DRS hops to reach each point when two RISs are installed at the marked location. 7 DRS are almost sufficient for covering the entire map.

Next, we obtain the achievable rate of each point in the map and compare the two scenarios by plotting heatmaps illustrating the achievable rate distribution which is shown in Fig. 5. Note that to obtain the rate at any point, we need to find the path of selected hops providing connectivity to that particular point and select the minimum rate between any two hops as the maximum achieved throughput at that point. Fig. 5 illustrates the improvement in terms of reachability and throughput that is achieved when only a single location is supplemented with RISs. This improvement arises from both the new virtual LOS paths and the increased signal power due to the beamforming gain.

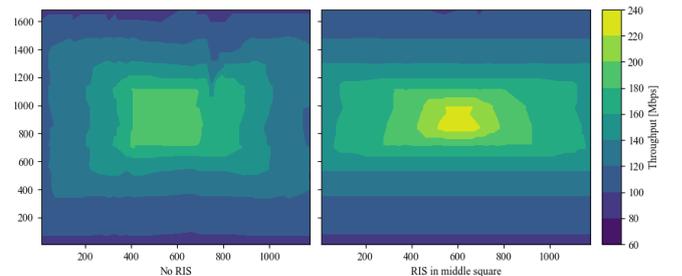

Fig. 5. Heatmap showing the distribution of achievable rate in the map. The left figure concerns the case where no RISs are installed, whereas the right figure shows the results when two RISs are installed in locations indicated in Fig. 4.

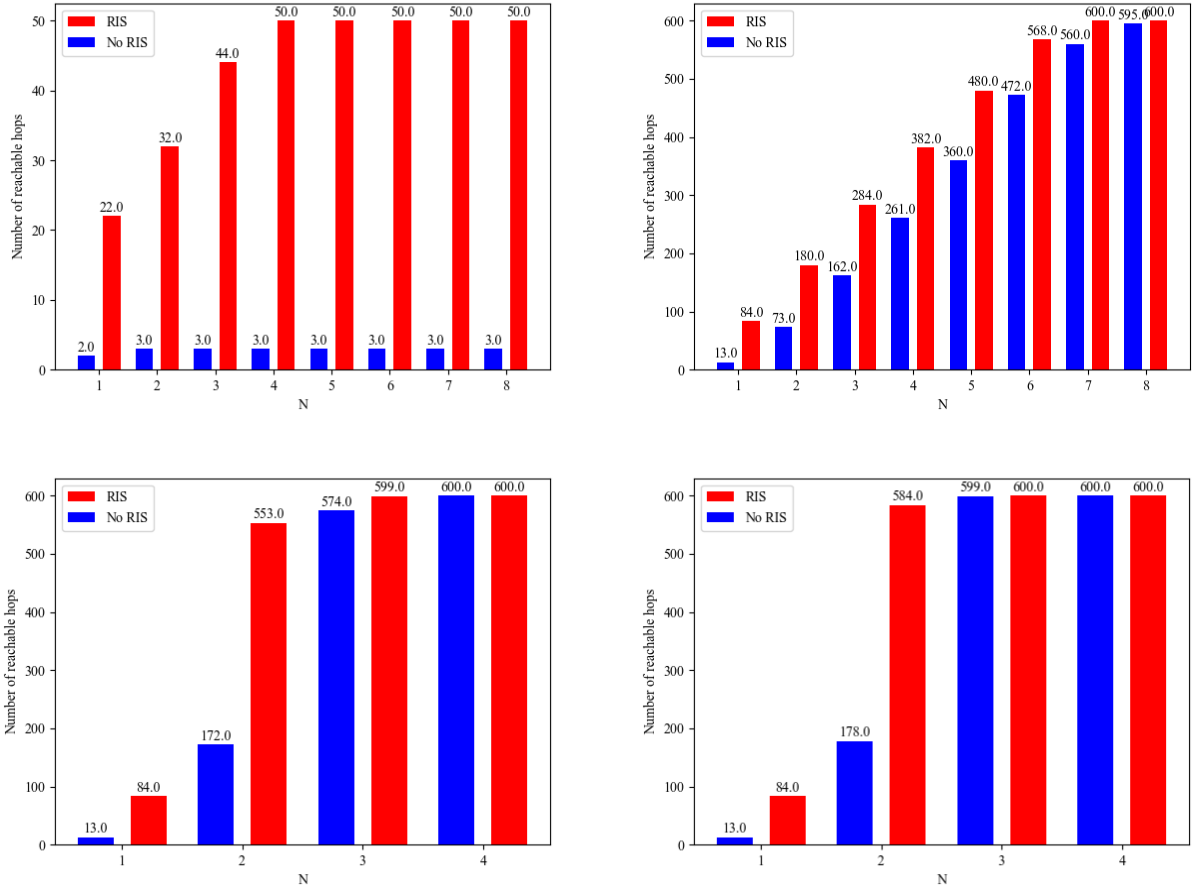

Fig. 6. The bars show the number of reachable points in the map for the SNR requirements of 41 dB (top left), 31 dB (top right), 21 dB (bottom left), and 11 dB (bottom right). For the stringent requirement of 41 dB, we observe that the reachability is limited by the distance between feasible points and, therefore, increasing the number of DRS hops does not increase the reachability.

Finally, Fig. 6 shows different bar plots considering different SNR requirements and illustrating the number of reachable points from the set of all feasible points shown in Fig. 2 that count to 600 points. For each SNR requirement, we increase the number of allowed DRS hops, $N$, successively up to a maximum of 8. In the case where an SNR requirement of 41 dB is assumed, we notice that without any RISs deployment we can only reach a maximum of 3 points because the distance, and therefore the path loss, between feasible points limits the reachability regardless of the number of DRS hops. Similarly, when deploying the RISs, and regardless of $N$, we can only reach a maximum of 50 points which is a great improvement in terms of reachability. For further illustration, we plot this specific scenario in Fig. 7 to better understand the reason for this behavior. We can observe that the distance between endpoints of long buildings edges incurs a path loss which prevents links satisfying the SNR requirement. In the case where RISs are installed, the beamforming gain allows the establishment of links through the RIS and to the other side of the long buildings. This is a clear limitation of our simplified selection of feasible points (i.e., building corners), to overcome this issue, the middle points of long buildings edges can be also added to the set of feasible points to increase the density of points and therefore allow less path loss values when selecting successive hops. Otherwise, more RISs can be deployed in appropriate location to stretch the reachability further. As for the other figures with higher SNR requirements, we can observe that all points can be reached by increasing $N$ and at a faster rate in the case where RISs are installed.

## V. NEXT STEPS

So far in this work we have investigated only the coverage problem for RIS and UAV assisted multi-hop backhauling. While providing connectivity and required backhaul capacity to a significant area is a crucial task, other important factors can be also accounted for. Energy efficiency factors can be considered aiming at reducing the number of required DRSs to provide the required capacity. Furthermore, reducing the transmission latency for the considered backhaul link, where each additional DRS introduces processing delay related to the decode-and-forward procedure, is of importance. Thus,

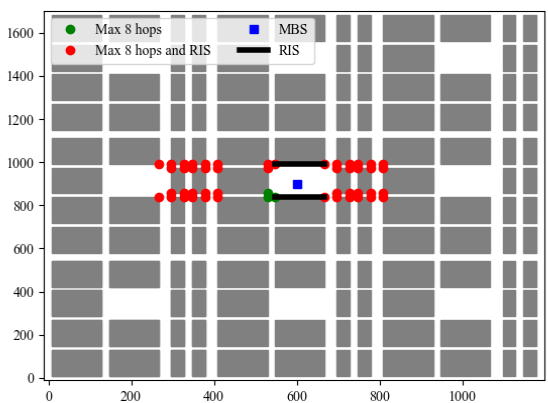

Fig. 7. Reachable points with and without RISs involvement for an SNR requirement of 41 dB. Notice that increasing N does not increase reachability.

among the envisaged joint optimization criteria one can consider the following:

- achieved throughput - the total data rate achieved in a multi-hop link, depending on the propagation conditions of each hop,
- energy consumption - the amount of energy required for deployment and operation of the DRSs involved in creation of the multi-hop link, including also the power consumption related to signal transmission (frontend processing),
- introduced latency - the delay in information transfer, taking into account both the propagation delays and the processing delays related to the need of additional signal processing performed at DRSs (signal decoding, reallocation of resources if needed).

Furthermore, we can also consider different, more complicated configurations of availability of RISs. So far, we have considered only fixed RIS mounting on buildings' facades. With the next steps we will investigate the positioning of DRSs knowing the availability and deployment configuration of RISs, considering the possible changes in their location. One can consider the use of mobile RISs, that can be mounted e.g. on vehicles or UAVs, thus providing more degrees of freedom with their deployment. Such a scenario will require advanced tools for optimization of the path selection and DRSs placement, thus both conventional optimization and machine learning tools will be accounted for in the investigation. Furthermore, use of additional context information that can be stored in databases, such as e.g., the Radio Environment Maps (REMs), will be considered to further improve the performance of the developed optimization algorithms. Finally, when in-band IAB configuration is considered, the problem needs to be extended to a general resource allocation problem with significant new types of interference, where the same time-frequency resources are allocated to end users or backhaul links.

## VI. Conclusion

In this work we study the idea of providing a multi-hop backhaul link between a macro BS and an UAV-mounted hotspot with the use of drone relay stations and reconfigurable intelligent surfaces. With this contribution we aim to optimize the coverage of the network by maximizing the number of available deployment locations for the drone access point, assuming minimum throughput constraints on the multi-hop link. We show that with the use RISs it is possible to increase the coverage by extending the individual links between macro BS and UAV or between two UAVs. Furthermore, we propose some improvements and open topics for further study in this area that will be conducted as the next steps of our research.

## Acknowledgment

The work has been realized within research project no. 2021/43/B/ST7/01365 funded by National Science Center in Poland.

## References

[1] Jaber, Mona, et al. "5G backhaul challenges and emerging research directions: A survey." IEEE access 4 (2016): 1743-1766
[2] P. Fabian, G. Z. Papadopoulos, P. Savelli and B. Cousin, "Performance Evaluation of Integrated Access and Backhaul in 5G Networks," 2021 IEEE Conference on Standards for Communications and Networking (CSCN), Thessaloniki, Greece, 2021, pp. 88-93, doi: 10.1109/CSCN53733.2021.9686110.
[3] U. Challita and W. Saad, "Network formation in the sky: unmanned aerial vehicles for multi-hop wireless backhauling," in Proc. IEEE Globecom, Dec. 2017.
[4] Zhang, X.;Wang, J.; Poor, H.V. "Joint Optimization of IRS and UAV-Trajectory: For Supporting Statistical Delay and Error-Rate Bounded QoS Over mURLLC-Driven 6G Mobile Wireless Networks Using FBC". IEEE Veh. Technol. Mag. 2022, 17, 55–63.
[5] Y. Liu et al., "Reconfigurable Intelligent Surfaces: Principles and Opportunities," in IEEE Communications Surveys & Tutorials, vol. 23, no. 3, pp. 1546-1577, thirdquarter 2021.
[6] S.A.H. Mohsan, et al., "Intelligent Reflecting Surfaces Assisted UAV Communications for Massive Networks: Current Trends, Challenges, and Research Directions". Sensors 2022, 22, 5278.
[7] S. Janji, A. Samorzewski, M. Wasilewska and A. Kliks, "On the Placement and Sustainability of Drone FSO Backhaul Relays," in IEEE Wireless Communications Letters, vol. 11, no. 8, pp. 1723-1727, Aug. 2022.
[8] A. Fouda, A. S. Ibrahim, I. Guvenc and M. Ghosh, "UAV-Based In-Band Integrated Access and Backhaul for 5G Communications," 2018 IEEE 88th Vehicular Technology Conference (VTC-Fall), Chicago, IL, USA, 2018, pp. 1-5, doi: 10.1109/VTCFall.2018.8690860.
[9] M. A. Abdel-Malek, A. S. Ibrahim, M. Mokhtar, K. Akkaya, "UAV positioning for out-of-band integrated access and backhaul millimeter wave network," Physical Communication, Vol. 35, 2019
[10] C. Pan, J. Yi, C. Yin, J. Yu and X. Li, "Joint 3D UAV Placement and Resource Allocation in Software-Defined Cellular Networks With Wireless Backhaul," in IEEE Access, vol. 7, pp. 104279-104293, 2019, doi: 10.1109/ACCESS.2019.2927521.
[11] Z. Wang, F. Zhou, Y. Wang and Q. Wu, "Joint 3D trajectory and resource optimization for a UAV relay-assisted cognitive radio network," in China Communications, vol. 18, no. 6, pp. 184-200, June 2021, doi: 10.23919/JCC.2021.06.015.
[12] P. Li and J. Xu, "UAV-Enabled Cellular Networks with Multi-Hop Backhauls: Placement optimization and Wireless Resource Allocation," 2018 IEEE International Conference on Communication Systems (ICCS), Chengdu, China, 2018, pp. 110-114, doi: 10.1109/ICCS.2018.8689218.
[13] U. Challita and W. Saad, "Network Formation in the Sky: Unmanned Aerial Vehicles for Multi-Hop Wireless Backhauling," GLOBECOM 2017 - 2017 IEEE Global Communications Conference, Singapore, 2017, pp. 1-6, doi: 10.1109/GLOCOM.2017.8254715.
[14] H. -B. Jeon, S. -H. Park, J. Park, K. Huang and C. -B. Chae, "RIS-assisted Aerial Backhaul System for UAV-BSs: An Energy-efficiency Perspective," 2021 IEEE Global Communications Conference (GLOBECOM), Madrid, Spain, 2021, pp. 1-6, doi: 10.1109/GLOBECOM46510.2021.9685565.
[15] E.M. Mohamed, M. Alnakhli, S. Hashima, M. Abdel-Nasser, "Distribution of Multi MmWave UAV Mounted RIS Using Budget Constraint Multi-Player MAB," Electronics, vol .12, 2023.
[16] P. Agyapong et al., "ICT-317669-METIS/D6.1 Simulation Guidelines", ICT-317669-METIS, Oct. 2013
[17] T. S. Rappaport, F. Gutierrez, E. Ben-Dor, J. N. Murdock, Y. Qiao and J. I. Tamir, "Broadband Millimeter-Wave Propagation Measurements and Models Using Adaptive-Beam Antennas for Outdoor Urban Cellular Communications," in IEEE Transactions on Antennas and Propagation, vol. 61, no. 4, pp. 1850-1859, April 2013, doi: 10.1109/TAP.2012.2235056.
[18] M. A. Kishk and M. -S. Alouini, "Exploiting Randomly Located Blockages for Large-Scale Deployment of Intelligent Surfaces," in IEEE Journal on Selected Areas in Communications, vol. 39, no. 4, pp. 1043-1056, April 2021, doi: 10.1109/JSAC.2020.3018808.
[19] P. Mogensen et al., "LTE Capacity Compared to the Shannon Bound," 2007 IEEE 65th Vehicular Technology Conference - VTC2007-Spring, Dublin, Ireland, 2007, pp. 1234-1238, doi: 10.1109/VETECS.2007.260.
[20] 3GPP, 5G, NR User Equipment (UE) Radio Access Capabilities, Document 3GPP TS38.306 v.17.2.0, Release 17, 2022.